\documentclass[aps,prl,floatfix,superscriptaddress,twocolumn]{revtex4}
\usepackage{graphics,bm,amsmath,amssymb,natbib,url,epsfig,psfrag, color,xcolor}
\usepackage{ulem}
\usepackage{comment}
\newcommand{\Renyi}[0]{R\'{e}nyi~}
\newcommand{\be}{\begin{equation}}
\newcommand{\ee}{\end{equation}}
\newcommand{\smk}{S^{(\alpha)}}
\newcommand{\bea}{\begin{eqnarray}}
\newcommand{\M}{M}
\newcommand{\eea}{\end{eqnarray}}

\begin{document}
\newcommand{\Tr}{\text{Tr}}
\title{Realistic protocol to measure entanglement at finite temperatures}
%How to measure 
%%number 
%entanglement 
%%entropy 
%at finite temperatures?}
%\title{Thermodynamic relations between number entanglement  and charge correlations}
%Equilibrium relations between number entanglement and occupation noise

%Measuring\Renyi Entropy change using occupation noise
\author{Cheolhee Han}
\affiliation{School of Physics and Astronomy, Tel Aviv University, Tel Aviv 6997801, Israel}

\author{Yigal Meir}
\affiliation{Department of Physics, Ben-Gurion University of the Negev, Beer-Sheva, 84105 Israel}
\affiliation{Department of Physics, Princeton University, Princeton, NJ 08540}
\author{Eran Sela}
\affiliation{School of Physics and Astronomy, Tel Aviv University, Tel Aviv 6997801, Israel}
\begin{abstract}
It is desirable to relate entanglement  of many-body systems to measurable observables. In systems with a conserved charge, it was recently shown that the number entanglement entropy (NEE) - i.e. the entropy change due to an unselective subsystem charge measurement - is an entanglement monotone. 
Here we derive finite-temperature equilibrium relations between \Renyi moments of the NEE, and multi-point charge correlations. These relations are exemplified  in   quantum dot systems where the desired charge correlations can be measured via a nearby quantum point contact. In quantum dots recently realizing the multi-channel Kondo effect we show that the NEE has a nontrivial universal temperature dependence which is now accessible using the proposed methods.
\end{abstract}

\maketitle
%\paragraph*{Introduction.}
Entanglement is a key concept in quantum mechanics, and its measures are being repeatedly used in many areas of physics, including basic quantum theory, quantum information and many-body physics. For systems at zero temperature or those described by a pure state, one can quantify the entanglement by the system's entanglement entropy. However, the task of measuring entanglement entropy in a many-body system is daunting as, in principle, one needs to measure the full density matrix, requiring access to all degrees of freedom (see, however, Ref.~\cite{Bartkiewicz2016priority, zhou2021direct}). %and a
As a result only very few experiments have been able to measure the entanglement entropy in specific systems of cold atomic Bose-Hubbard chains~\cite{islam2015measuring, lukin2019probing} and trapped-ion quantum simulators~\cite{brydges2019probing, elben2020mixed}. Thus the measurement of entanglement entropy remains one of the outstanding challenges in many-body physics. The task of experimental quantification of entanglement at finite temperatures is even more challenging,  as the standard entanglement entropy measure is no longer uniquely defined in this case.

An important step forward was the recognition \cite{samuelsson2003orbital,beenakker2003proposal,klich2006measuring,klich2009quantum,song2010general,song2011entanglement,rachel2012detecting,zhang2022measuring}
of a relationship between entanglement entropy 
and fluctuations in condensed matter systems. For example, for a system with a conserved total number of particles $N=N_A+N_B$ a ground state with fluctuating number of particles in  subsystem $A$ directly implies a nonlocal quantum correlation with subsystem $B$.
These works %has been recognized 
%in recent years 
%in various works
focused mainly on generating~\cite{samuelsson2003orbital,beenakker2003proposal}, measuring~\cite{klich2006measuring,klich2009quantum}, and understanding of the scaling of many-body quantum entanglement~\cite{song2010general,song2011entanglement,rachel2012detecting}. In fact, for  non-interacting fermions an exact relation between current moments and entanglement entropy was found~\cite{klich2009quantum}. %(
Note, however, that in general, such %number or 
current correlations do not fully capture  the charge-neutral contributions~\cite{klich2006measuring}, often referred to as configuration entropy~\cite{wiseman2003entanglement,barghathi2019operationally, goldstein2018symmetry,xavier2018equipartition,bonsignori2019symmetry}.% and thus the number entanglement entropy  deduced from charge or current fluctuations only yields a lower bound for the entanglement entropy~\cite{klich2006measuring}). 

Yet, while all these previous works considered pure states, general condensed matter systems are described by mixed states, e.g. due to finite temperature.  Then, as mentioned above, 
the entanglement entropy
%, a standard entanglement measure for pure states }%Then the standard definition of entanglement entropy 
has to be replaced by generalized entanglement measures, such as negativity~\cite{peres1996separability,vidal2002computable} or number entanglement entropy (NEE)~\cite{macieszczak2019coherence,ma2021symmetric}. However, unlike the case of a pure state, mentioned above, relations between mixed state entanglement quantifiers 
and charge/current correlations are not known, which makes the task of measuring %the 
entanglement at finite temperatures %measure 
even harder. Previous successful attempts to quantify entanglement for mixed states relied on random unitary measurements~\cite{elben2020mixed,zhou2020single,neven2021symmetry,vitale2022symmetry,rath2022entanglement},
%gray2018machine
which are hard to apply in general condensed matter systems. Here, we provide exact thermodynamic relations between moments of the NEE  and measurable subsystem charge fluctuations. Consequently, measuring these quantities will facilitate obtaining information about entanglement in many-body condensed matter systems at finite temperatures.
 
The NEE $\Delta S$  is defined as the  entropy change of a  density matrix $\rho$ upon an unselective measurement of $N_A$~\cite{ma2021symmetric}, 
\begin{align}
\label{def:DeltaSm}
\Delta S=- \Tr[\rho_{\hat{N}_A}\log  \rho_{\hat{N}_A}]+\Tr[\rho\log\rho],
\end{align}
%\ES{If we adopt this notation change we will change it everywhere $\Delta S_m \to \Delta S_N$, $\rho_m \to \rho_N$}
where $\rho_{\hat{N}_A}=\sum_{N_A} \Pi_{N_A}\rho\Pi_{N_A}$ and $\Pi_{N_A}$ is a projector to a subspace with subsystem charge $N_A$. A similar quantity in terms of the reduced density matrix was studied in Ref.~\onlinecite{ares2022entanglement}. Equivalently, $\rho_{\hat{N}_A}$ is obtained from $\rho$ by annihilating all off-diagonal matrix elements with respect to $N_A$. 
When $N_A$ uniquely specifies the subsystem's state, $\Delta S$ coincides with the entanglement entropy for a pure state (one can easily generalize NEE to include spin and other globally conserved quantum numbers). Note that even for a mixed state,  $\Delta S$ does not depend on which of the subsystems one traces over, unlike the standard entanglement entropy.

In fact, our definition in Eq.~(\ref{def:DeltaSm}) is a %special 
quantifier of coherence~\cite{baumgratz2014quantifying}, associated with a conserved quantity~\cite{macieszczak2019coherence}. As the entanglement entropy %also 
reflects the coherence in the system, the NEE measures charge coherence and cannot be simply extracted from the charge distribution $P(N_A)$. Instead, as we show below, one has to measure charge correlation functions in order to quantify the NEE in the system. 

The difficult task of directly measuring the entanglement entropy or negativity in many-body systems was circumvented in the literature by suggestions to measure instead its first \Renyi moments~\cite{abanin2012measuring,daley2012measuring,pichler2013thermal,gray2018machine,elben2018renyi,elben2020mixed}. We will apply a similar strategy for the NEE. The \Renyi moments of $\Delta S$ are defined by %We define another witness of nonseparability,  the change of $k$th order\Renyi entropy (RE),
\begin{align}
\Delta \smk=\frac{1}{1-\alpha}\log\Big(\frac{\Tr[\rho_{\hat{N}_A}^\alpha]}{\Tr[\rho^\alpha]}\Big).\label{eq:Renyi_def}
\end{align}
While the \Renyi moments of the NEE do not inherit its monotonicity, for general mixed states, %where
$(\Tr[\rho^\alpha] \ne 1)$, it can be shown that $\Delta \smk > 0$ is a sufficient condition for quantum entanglement or quantum coherence~\cite{SM}. 
One can formally extract  the NEE from its moments (see also Refs.~\cite{d2021alternative,carisch2022tensor}) via   $\Delta S  = \lim_{\alpha \to 1} \Delta \smk$  \footnote{Note, however, that one has to take into account that $\Delta S^{(\alpha)}$ may be a non-monotonic function of $\alpha$, different from the \Renyi moments of entropy~\cite{zyczkowski2003renyi}.}. Hereafter we write the NEE as $\Delta S^{(1)}$ for convenience.
For pure states $\Tr[\rho^\alpha]=1$ and $\Delta \smk=\frac{1}{1-\alpha} \log \sum_{N_A} (P(N_A))^\alpha$  becomes the \Renyi entropy of the subsystem charge distribution $P(N_A)$. 
As mentioned above, for general mixed states, the NEE and its \Renyi moments cannot be obtained from $P(N_A)$ alone (see examples in~\cite{SM}). %For general mixed states, the NEE (and its \Renyi moments) differs from the corresponding entropy of the subsystem charge distribution, referred to as number entropy. We display this explicitly in specific examples~\cite{SM}, where the former vanishes at high temperatures despite that the latter increases.
When the subsystem charge $N_A$ uniquely specifies its state, $\Delta \smk$ coincides with the usual \Renyi entropy.   %The second \Renyi moment of the NEE can be used, for example, to construct a computable mixed-state entanglement monotone that reduces to known measures of  entanglement in the pure state limit~\cite{carisch2022tensor}.

Typically, \Renyi moments are obtained using multiple copy methods~\cite{abanin2012measuring,daley2012measuring,pichler2013thermal,islam2015measuring,gray2018machine,cornfeld2018entanglement,cornfeld2019measuring,daniel2020identification}, which are quite difficult to generate experimentally, especially in condensed matter systems. Here, focusing on general thermal states, rather than physically implementing $\alpha$ copies of the system,  we demonstrate that the $\alpha-$th \Renyi moment of the NEE of a thermal state at temperature $T$ is directly related to a correlation function of the same system at temperature $T/\alpha$.

Substituting $\rho(T)=Z(T)^{-1}\sum_i  e^{-E_i/T} |i \rangle \langle i|$ where $Z(T)=\sum_{i}e^{-E_i/T}$, and $|i \rangle$ are the eigenstates of $H$, $H|i\rangle=E_i |i\rangle$,  for all the factors of $\rho$ in the the numerator in Eq.~\eqref{eq:Renyi_def},
we find
\begin{align}
\Tr[(\rho_{\hat{N}_A})^\alpha]&=\sum_{N_A} \Tr[(\rho\Pi_{N_A})^\alpha]=\sum_{j_1,...j_\alpha}\frac{e^{-(E_{j_1}+...+E_{j_\alpha})/T}}{Z(T)^\alpha}\nonumber\\
\times\langle & j_1|\Pi_{N_A}|j_\alpha\rangle\langle j_\alpha|\Pi_{N_A}|j_{\alpha-1}\rangle...\langle j_2|\Pi_{N_A}|j_1\rangle \label{eq:rhom_k}.
\end{align}
We now note that this expression 
is identical up to a multiplicative factor of $Z(T') / Z(T)^{\alpha}$ to
%is equivalent to 
the $\alpha-$point imaginary time Green's function of projection operators,% $\Pi_{N_A}(\tau)$, 
\begin{align}
&\langle \Pi_{N_A}(-i\tau_{\alpha-1})\Pi_{N_A}(-i\tau_{\alpha-2})\cdots \Pi_{N_A}(-i\tau_1)\Pi_{N_A}(0)\rangle_{T'} \nonumber\\
=&\sum_{j_1,...j_\alpha}\frac{e^{-E_{j_1}/T'}}{Z(T')}\langle j_1|\Pi_{N_A}|j_\alpha\rangle\cdots\langle j_2|\Pi_{N_A}|j_1\rangle \label{eq:ppcor_imag}\\
&\times e^{(E_{j_1}-E_{j_\alpha})\tau_{\alpha-1}} e^{(E_{j_\alpha}-E_{j_{\alpha-1}})\tau_{\alpha-2}}\cdots e^{(E_{j_3}-E_{j_2})\tau_1},\nonumber
\end{align}
evaluated at temperature $T' = T/\alpha$ and imaginary times $\tau_j = j/T$.
%by setting the temperature to $T'=T/\alpha$ and $\tau_j = j/T$. 
We also note that the denominator of Eq.~\eqref{eq:Renyi_def} can be written as $\Tr[\rho^\alpha]=Z(T/\alpha)/Z(T)^\alpha$. Combining these results, we obtain
%Combining this with the denominator of Eq.~(\ref{eq:Renyi_def}) we obtain
%Eq.~\eqref{eq:central_general}.
 %Our underlying result is expressing $\Delta \smk$ %, the $\alpha-$th \Renyi NNE moment,   in general thermal states as a $\alpha-$point imaginary time correlation function of  projection operators  $\Pi_{N_A}(\tau) =e^{i H \tau }  \Pi_{N_A}  e^{- iH \tau }$ at temperature $T/k$ %(see derivation below)
\begin{align}
\Delta \smk=\frac{1}{1-\alpha}\log\bigg[\sum_{N_A}&\Big\langle\Pi_{N_A}\Big(\frac{\alpha-1}{iT}\Big)\Pi_{N_A}\Big(\frac{\alpha-2}{iT}\Big)\cdots \nonumber\\
\times&\Pi_{N_A}\Big(\frac{1}{iT}\Big)\Pi_{N_A}(0)\Big\rangle_{T/\alpha} \bigg].\label{eq:central_general}
\end{align}
This equation, which is the main result of this paper, expresses the \Renyi  NEE at temperature $T$ in terms of a correlation function at temperature $T/\alpha$. 
%This embodies a realization of a multi copy replica approach to measure the \Renyi entropy for thermal states. %This mapping relies on thermal equilibrium. 

In order to demonstrate the power of this relation, let us focus on a many body system in the Coulomb blockade regime, 
%When the subsystem supports only two charge states, 
such as a quantum dot (QD) or a metallic grain with a large charging energy. %the correlation function becomes that of the subsystem charge $N_A$ itself. W
In this case, the subsystem supports only two charge states, and
we provide an  explicit protocol to extract the 2nd and 3rd \Renyi moments solely from the $2-$point charge correlation function of the QD. %\YM{\sout{ We discuss how the resulting information can be extracted from the current noise of a charge detector attached to the QD.}}
These relations allow to measure an entanglement quantifier in mesoscopic systems at finite temperatures.%\sout{thermal equilibrium}}.

%The projectors $\Pi_{N_A}$ are physical observables constructed from the subsystem charge operator $\hat{N}_A$.  Next, we relate the correlation function in Eq.~(\ref{eq:central_general})  to a subsystem charge correlator in cases where subsystem $A$ or $B$ admits only two charge states. 

%\paragraph*{Two charge states.}
Consider a subsystem $A$ with only two charge states, denoted for %simplicity 
$N_A=0,1$. 
%This describes for example a QD which is either occupied or empty. %There is no restriction on subsystem $B$ which can be, e.g., another QD or a lead as we will discuss in the examples in Sec.~\ref{se:examples}. 
In this case the projection operators become linear in $N_A$, $\Pi_{N_A=1}=\hat{N}_A$,  $\Pi_{N_A=0}=1-\hat{N}_A$. Thus Eq.~(\ref{eq:central_general}) becomes a sum of $q-$point ($q\leq \alpha$) charge correlators.

\begin{figure*}[]
\includegraphics[width=\textwidth]{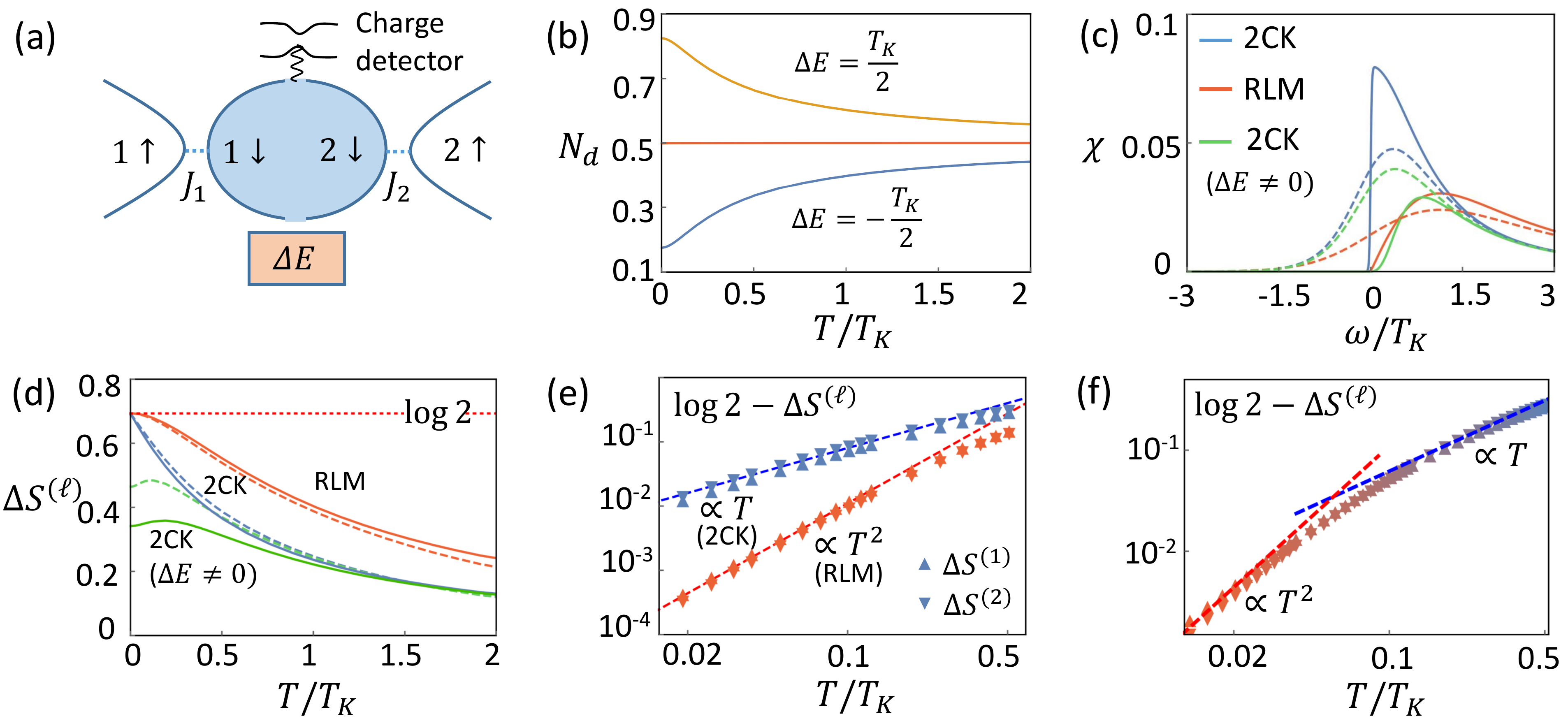}
\caption{(a) The charge-multi-channel-Kondo setup \cite{matveev1995coulomb,furusaki1995theory,iftikhar2015two,iftikhar2018tunable}, shown here for two channels (2CK). In each of the channels ($j=1,2$) an electron can hop from the metallic dot to its respective lead with matrix elements $J_j$ and
thus %by flipping its spin and 
change the dot occupancy by unity, $N_d=0 \leftrightarrow 1$, effectively flipping the ``impurity spin".  The gate detuning from charge degeneracy $\Delta E$ serves as an effective magnetic field. The asymmetry $T_-\propto(J_1-J_2)^2$ allows to study the crossover between two and one-channel Kondo models. %We added a
An additional quantum point contact coupled electrostatically to the dot %, which 
allows measurements of the dot charge and its fluctuations.  (b) The charge $N_d(T)$ and (c) its fluctuations $\chi(\omega)$, required for the evaluation of the \Renyi moments $\Delta S^{(\alpha)}$ of the number entanglement entropy (NEE). $N_d(T)$ is shown for $\Delta E= T_K/2$ (yellow), $0$ (red), $- T_K/2$ (blue) with $T_-=0$.  $\chi(\omega)$ is shown for $T=0.01T_K$ (solid lines) and $T=0.5 T_K$ (dashed line) for $T_-=\Delta E=0$ (2CK), $T_- =T_K$ (resonant level model - RLM) with $\Delta E=0$, and  for $\Delta E=T_K/2$ with $T_-=0$. In the plot, $\delta$ function peaks~\cite{SM} are not drawn. (d) The \Renyi moment of NEE $\Delta S^{(2)}$ obtained from $N_d$ and $\chi(\omega)$ using Eq.~\eqref{eq:central_Renyi23} (solid lines). The NEE $\Delta S^{(1)}$ is also shown for comparison (dashed lines). We compare the 2CK case, the RLM case, and also the 2CK with finite level detuning $\Delta E$.  (e) $\log-\log$ plot of $\log(2)-\Delta S^{(\alpha)}(T)$ for the 2CK and RLM cases, demonstrating distinct %low temperature 
power law behaviour; up (down) triangles refer to $\alpha=1$ ($\alpha=2$). (f) $\log(2)-\Delta S^{(\alpha)}(T)$ for a small channel anisotropy scale $T_- / T_K =  0.1$, displaying a 1CK-2CK crossover.% from the 2CK behavior for $T>T_-$ (blue) to 1CK %low temperature 
%behavior for $T<T_-$ (red). 
%(a) Multi channel charge Kondo setup with an additional QPC attached to the QD allowing to measure its charge and its charge fluctuations. 
%(b) $\chi(\omega,T)$ for $T=0.01T_K$ (solid lines) and $T=0.5 T_K$ (dashed line) for $T_-=\Delta E=0$, $T_- =T_K$ with $\Delta E=0$, and $\Delta E=T_K/2$ with $T_-=0$. In the plot, $\delta$ function peaks are not drawn.
%(c) $\Delta S_m$ and $\Delta S_m^{(2)}$ for $T_-=0$  and $T_-=T_K$ with $\Delta E=0$. 
%(d)$\log-\log$ plot of $\log(2)-\Delta S_m^{(\alpha)}(T,T_-)$. Circles (squares) are $\alpha=1$ ($\alpha=2$) (e) Occupation number of the metallic dot for $\Delta E=T_K/2,0,-T_K/2$ from uppermost to lowermost. (f) $\Delta S_m^{(1)}$ (solid lines) and $\Delta S_m^{(2)}$ (dashed lines) versus $T$ for $\Delta E=0$ (upper) and $\Delta E=T_K/2$ (lower).  
}
\label{fig:2ckresult}
\end{figure*}

Notably, the 2nd and 3rd \Renyi moments of the NEE in Eq.~\eqref{eq:central_general} can be written explicitly in terms of $\langle \hat{N}_A\rangle_{T/\alpha}$, the occupation number of $A$ and $\langle \hat{N}_A(-i(\alpha-1)/T)\hat{N}_A(0)\rangle_{T/\alpha}$, the 2-point imaginary time charge correlator.
%\subsection{From imaginary to real time}\label{se:imaginaryToRealtime}
Furthermore, at thermal equilibrium the imaginary time correlator $\langle \hat{N}_A(-i\tau)\hat{N}_A(0)\rangle_T$ can be related with the %frequency domain
Fourier transform of the %  appropriate
real-time correlator, %. The charge noise $\chi(\omega,T)$ in 
defined as
\begin{align}
&\chi(\omega,T)=\int dt e^{i\omega t}\langle \hat{N}_A(t)\hat{N}_A(0)\rangle_{T}.
\end{align}
This is the charge noise of the QD.
%Comparing with the imaginary time Green's function in Eq.~\eqref{eq:nncor_imag}, we obtain the relation $\langle \hat{N}_A(-i\tau)\hat{N}_A(0)\rangle_T=\int \frac{d\omega}{2\pi} \chi(\omega,T) e^{-\omega\tau}$. 
%Therefore $\Lambda(\tau,T)$ is obtained from $\chi(\omega,T)$. After putting specific $\tau$ and $T$ as in Eq.~\eqref{eq:nncor_imag_half}, we obtain the change of the 2nd and 3rd RE %shown in Eqs.~\eqref{eq:central_Renyi23}. Using this way, in principle we can relate $k$-point imaginary time function (see Eq.~\eqref{eq:central_general}) and $k$-point real time correlator. 
As a result Eq.~(\ref{eq:central_general}) becomes~\cite{SM}
\begin{equation}
\begin{split}
\Delta S^{(2)}=&-\log\Big[1-2\langle\hat{N}_A\rangle_{T/2}+2\int\frac{d\omega}{2\pi}\chi\big(\omega,T/2\big) e^{-\frac{\omega}{T}}\Big],\\
\Delta S^{(3)}=&-\frac{1}{2}\log\Big[1-3\langle \hat{N}_A\rangle_{T/3}+3\int \frac{d\omega}{2\pi}\chi\big(\omega,T/3)e^{-\frac{2\omega}{T}}\Big].
\end{split}\label{eq:central_Renyi23}
\end{equation}
This is our main result for Coulomb blockaded systems. It allows one to measure \Renyi moments of the NEE in thermal states at temperature $T$, from charge correlations measured at temperature $T/\alpha$.  The advantage of our approach is that these charge correlations have been repeatedly measured in mesoscopic systems  using charge sensing techniques (see, e.g.,~\cite{fujisawa2006bidirectional,gustavsson2006counting,gustavsson2006countingL,dial2013charge,wilen2021correlated}). In the supplementary material ~\cite{SM} we relate explicitly  $\chi(\omega,T)$ to the voltage-dependent noise~\cite{blanter2000shot} in a quantum-point-contact charge detector electrostatically coupled to the QD as depicted in Fig.~\ref{fig:2ckresult}(a). 

When subsystem $A$ fluctuates between more charge states, the projectors $\Pi_{N_A}$ become nonlinear functions of the charge operator. For example for three charge states denoted $N_A=0,1,2$ we have $\Pi_{N_A=0}=\frac{1}{2}(\hat{N}_A-1)(\hat{N}_A-2)$, and so on. %, $\Pi_{N_A=1}=\frac{1}{2}\hat{N}_A(\hat{N}_A-2)$, and $\Pi_{N_A=2}=$. 
Then in order to measure the NEE, Eq.~\eqref{eq:central_general} involves higher-point charge correlators. In the context of full counting statistics~\cite{matsuo2020full}, higher moments such as the third moment of current correlations $\langle  I(t_1) I(t_2) I(t_3) \rangle$ have been measured, see for example Refs.~\onlinecite{reulet2003environmental,bomze2005measurement}. Such techniques may allow to extend our results to the case with multiple-charge states, and similarly for higher \Renyi moments.

Eqs.~\eqref{eq:central_Renyi23} apply generally to Coulomb blockaded systems.
The simplest possible example is a spinless double dot system. The entanglement of $\log 2$ due to coherent hopping of an electron between the two dots can be directly computed from the density matrix, or equivalently from the charge-charge correlation via Eq.~\eqref{eq:central_Renyi23}, as demonstrated in the supplementary material~\cite{SM}. However, the power of the relations \eqref{eq:central_Renyi23} lies in the prospect of applying them to strongly correlated systems, such as magnetic impurities embedded in a continuum, which can lead to a multitude of Kondo effects and to a competition between spin correlations and screening in multiple-impurity systems (for a review see, e.g., \cite{grobis2007kondo}).
Here we exemplify the usefulness of these relations for multi-channel Kondo systems.

The multi-channel Kondo (MCK) model describes an impurity spin-1/2 interacting antiferromagnetically with $\M$ spinfull channels with continuous density of states. For $\M=1$ the impurity is perfectly screened at $T=0$, while for $\M>1$ the ground state is a non-Fermi liquid, due to overscreening of the impurity by the $\M$ channels~\cite{nozieres1980p}. In both cases, though,
the spin-screening cloud [see, e.g, \onlinecite{barzykin1998screening,affleck2009entanglement}] leads to a zero-temperature entanglement entropy of $\log2$. Recently, finite-temperature entanglement measures, such as the entanglement of formation (EoF)  and negativity (${\cal N}$), have been applied to this model ~\cite{lee2015macroscopic,kim2021universal}, demonstrating, for both quantities, a low-temperature scaling behavior of 
\be
 {\rm{EoF}}, {\cal N} \simeq \log 2 - a_{\M} T^{2\Delta_{\M}},
 \label{eq:EoF}
\ee
where $a_{\M}$ is an $\M$-dependent numerical coefficient and $\Delta_M$ is the scaling dimension of the impurity spin operator in the $\M$-channel problem. As determined by the conformal field theory solution~\cite{affleck1991critical}, $\Delta_\M=2/(2+\M)$ for $\M \ge 2$, and for the single channel case $\Delta_1=1$.  Unfortunately, these entanglement measures, while calculable, are practically impossible to measure (as they require measurements of all elements of the density matrix). Below we demonstrate that  the entanglement measure NEE for the charge-Kondo model, and its \Renyi moments, which can be directly measured by applying Eqs.~(\ref{eq:central_Renyi23}), obey the same scaling relation as Eq.~(\ref{eq:EoF}), and thus allow direct measurement of the finite-temperature entanglement and the scaling dimensions of MCK systems.

Unlike the commonly observed one-channel Kondo effect, signatures of a two-channel spin-Kondo behavior in quantum dots have only been reported in \cite{potok2007observation,keller2015universal}. A major step forward has been recently taken in Refs.~\onlinecite{iftikhar2015two,iftikhar2018tunable} which utilized a mapping between the two charge states of a metallic dot $N_d=0,1$ and spin $S_z=\pm 1/2$ \cite{matveev1995coulomb,furusaki1995theory} to demonstrate scaling and crossover between single and multichannel $\M=1,2,3$ Kondo effects [here the number of channels is controlled by the number of one dimensional leads attached to the dot, see Fig.~\ref{fig:2ckresult}(a)]. Since, in this case of a charge-Kondo effect, the role of spin is played by the deviation of the occupation from half-filling, then charge correlations in the charge-Kondo system map onto spin correlation in spin-Kondo system, thus allowing us to extract the finite temperature entanglement measure  NEE for non-Fermi liquid MCK systems by utilizing Eqs.~(\ref{eq:central_Renyi23}). In this case the NEE reduces at $T=0$ to the full entanglement entropy.

To be specific, the charge-MCK Hamiltonian is given by \cite{matveev1995coulomb,furusaki1995theory}
\begin{align}
    H=&\sum_{j=1}^M J_{j} \sum_{k,k'}(c_{j\uparrow k}^\dagger c_{j\downarrow k'} S_- +c_{j\downarrow k}^\dagger c_{j\uparrow k'} S_+)\nonumber\\
    &+\sum_{j,k,\sigma}k c_{j\sigma k}^\dagger c_{j\sigma k}+\Delta E S_z.
    \label{eq:MCK}
\end{align}
The system is depicted in Fig.~\ref{fig:2ckresult}(a) for the case $M=2$. Here $J_{j}$ is the tunneling between the $j$-th lead and the dot,  $c_{j\sigma k}$ and  $c^\dagger_{j\sigma k}$ denote annihilation and creation operators, respectively, for electrons with  momentum $k$ in the leads for $\sigma=\uparrow$ or  in the dot for $\sigma=\downarrow$. Here the spin flip operator changes the charge state of the  dot, $S_+|0\rangle= |1\rangle$, and $S_z=(|1\rangle \langle 1|-|0\rangle \langle 0|)/2=\hat{N}_d-1/2$, where $|0\rangle$ and $|1\rangle$  are the empty and occupied dot states, and $\hat{N}_d$ is the dot
number operator. $\Delta E$ is a gate voltage detuning away from the charge  degeneracy point, which corresponds to a magnetic field in the conventional MCK model. The electrons are assumed to be spinless, e.g. due to a large magnetic field. The Hamiltonian (\ref{eq:MCK}) is identical to the spin-MCK Hamiltonian, where the role of the local spin is played by the occupation states of the metallic dot. 

 Consider now the NEE  between the QD and the leads. Using Eqs.~(\ref{eq:central_Renyi23}), which expresses the NEE as a correlation function, we find for the second \Renyi NEE
\be
    \Delta S^{(2)}=-\log\Big[\frac{1}{2}+2\langle S_z(\frac{1}{iT})S_z(0)\rangle_{T/2}\Big].
    \label{eq:sscorreltions}
\ee
%\ES{where the $z$-component of the impurity spin is simply the  ``charge" operator of the quantum dot, $S_z=N_A-1/2$.} 
The second term is a two-point correlation function of the impurity spin operator in imaginary time. Using the general conformal field theory 
 correlation function $\langle \mathcal{O}(-i\tau) \mathcal{O}(0) \rangle_T = \left(\frac{\pi T}{\sin \pi \tau T} \right)^{2 \Delta_{\mathcal{O}}}$ for an operator $\mathcal{O}$ with scaling dimension $\Delta_\mathcal{O}$, we obtain~\cite{SM}
%solution of the MCK model, it obeys at low temperatures a scaling form~\cite{SM}
\be
\label{szszDeltaM}
\langle S_z(\frac{1}{iT})S_z(0)\rangle_{T/2} \propto \left( \frac{\pi T/2}{\sin (\frac{\pi}{T} T/2)} \right)^{2\Delta_{M}} \sim  T^{2\Delta_{M}}.
\ee
Furthermore, we find that at low temperature the NEE $\Delta S^{(1)}$ is also dominated by the same two point correlator~\cite{SM}. 
%(see section VI in Ref.~\cite{SM}).
%If you give sectoins to the SM within the text, we should do this consistently

Accordingly the entanglement measure NEE and its \Renyi moments obey the exact same scaling behavior as EoF and  ${\mathcal{N}}$ in Eq.~(\ref{eq:EoF}). For example for the 1CK case we obtain a quadratic temperature dependence, $\Delta S^{(\alpha)}_{1CK} \simeq \log 2 - a_{1\alpha} T^{2}$, whereas in the 2CK case the dependence is linear, $\Delta S^{(\alpha)}_{2CK} \simeq \log 2 - a_{2\alpha} T$, where now the non-universal coefficients $a_{M\alpha}$ depend also on the \Renyi moment $\alpha$. As pointed out in Refs.~\cite{lee2015macroscopic,kim2021universal}, this linear behavior, corresponding to $\Delta_{2}=1/2$, is an indication of the existence of a  Majorana fermion in the ground state. 
%For $\M=3$ Eq.~(\ref{eq:sscorreltions}) yields a $T^{4/5}$ temperature dependence, which also coincides with that found for negativity~\cite{kim2021universal}.

\begin{comment}
Thus at zero temperature the second term involving the $S_z$ correlation  vanishes and we obtain a $\log 2$ entropy, independent of the number of channels.  The temperature correction is then expected to be
\be
%\label{eq:MCK}
\Delta S^{(2)} \simeq \log 2 - a T^{2\Delta_{M}},
\ee
exactly as in Eq.~(\ref{eq:MCK}), where $a$ is another numerical coefficient. In fact, we find~\cite{SM} this result holds %for $\Delta S_m^{(\alpha)}$ for 
any $k$ including $\alpha=1$.% (with a $k$ dependent coefficient $a_k$).
\end{comment}

The temperature dependence obtained in Eq.~(\ref{szszDeltaM}) is based on a general scaling argument. In order to make a more quantitative comparison with experiments for the case of the 1CK and 2CK models, we can use the exact solution of the corresponding low-temperature Hamiltonian. This allows us  to obtain results for the temperature behavior along the crossover between 1CK and 2CK behaviors. Also, the same model allows us to provide quantitative results for the simpler resonant level model.

After a renormalization process at low temperatures, % and E-K transformation, 
the Hamiltonian for the $M=2$ MCK model can be mapped into a Majorana resonant level model~\cite{emery1992mapping,zarand2000analytical}
\begin{equation}
\begin{split}
     H=&\sum_k k\gamma_{x,-k}\gamma_{x,k}+k\eta_{x-k}\eta_{x,k}\\
    &+i\sqrt{\frac{2T_K}{L}}\sum_k \gamma_k \eta_d +i\sqrt{\frac{2T_-}{L}}\sum_k \eta_k \gamma_d+i\Delta E\gamma_d\eta_d.   
\end{split}
\end{equation}
Here $T_K$ %
%=(J_1+J_2)^2/(16 a)$ with
%the density of state $\nu$, 
%WE DO NOT NEED TO SHOW THIS DEPENDENCE  - WHICH IS ACTUALLY DIFFERENT
is the Kondo energy scale and $T_-\propto (J_1-J_2)^2$~\cite{mitchell2016universality}
%/(16 a)$ 
is an energy scale associated with channel asymmetry. Here $\gamma_d$ and $\eta_d$ are local Majorana zero modes, and $\gamma_{x,k}$,$\eta_{x,k}$ represent the mode expansion of a Majorana fields. The charge occupation operator of the metallic dot is written in terms of the local Majorana operators as $\hat{N}_d=\frac{1}{2}+ i\gamma_d\eta_d \equiv \hat{N}_A$. 

This model describes different regimes. The 2CK state corresponds to $\Delta E=T_-=0$. As long as ${\rm{max}} \{ \Delta E, T_- \} \ll T_K$, this model describes the vicinity of the 2CK fixed point.
As $T_-$ increases, it faithfully describes the crossover~\cite{pustilnik2004quantum,mitchell2012universal} from the 2CK to the 1CK fixed point.
%emerging below the energy scale ${\rm{max}} \{ \Delta E, T_- \} $.
Additionally, for the special value $T_-=T_K$, this model maps into the non-interacting resonant level model (RLM), which can be realized in a single level spinless QD, with on site energy $\Delta E$ and width $\Gamma \equiv T_K$. %In the results below, we discuss the 2CK state and the RLM model in parallel.

This model can be solved exactly \cite{emery1992mapping}. The results for the occupation number $N_d$ and $\chi(\omega)$, which can be measured using the charge detector, were obtained by solving numerically the resulting integral expressions~\cite{SM}
%quations 
and are shown in Fig.~\ref{fig:2ckresult}(b,c)  for selected model parameters. %By 2CK we refer to $T_-=0$ and by RLM to $T_-=T_K$. For integral expressions see Ref.~\cite{SM}). 
For $\Delta E=0$ (either 2CK or RLM), by symmetry $N_d=1/2$, and the NEE \Renyi moments $\Delta S^{(\alpha)}$ can be extracted using Eq.~(\ref{eq:central_Renyi23}) solely from the charge noise  $\chi(\omega)$. We plot in Fig.~\ref{fig:2ckresult}(d) the resulting $\Delta S^{(2)}$, and for comparison~\cite{SM} the NEE ($=\Delta S^{(1)}$) which is an entanglement monotone~\cite{ma2021symmetric}, with very similar behavior. Specifically, the resulting $\Delta S^{(\alpha)}$ for the 2CK state is in agreement with our field theory scaling results as seen in the $\log-\log$ plot in Fig.~\ref{fig:2ckresult}(e). For the RLM the temperature scaling is quadratic. 

In Fig.~\ref{fig:2ckresult}(f) we show the crossover from 2CK to 1CK behavior which can be observed for a small channel anisotropy $T_-$. While  $\lim_{T \to 0} \Delta S^{(\alpha)}$ remains $\log 2$ for any $J_-$, since by symmetry $N_d=1/2$, the temperature dependence becomes quadratic $\log 2-\Delta S^{(\alpha)}\propto T^2$ for $T < T_-$. % This suggests that the entanglement is more fragile in the 2CK case compared to the 1CK case.
For a finite gate voltage detuning from the fixed point ($\Delta E\ne 0$) we have that $\langle N_d\rangle\ne 1/2$ as seen in Fig.~\ref{fig:2ckresult}(b). As a result, $\lim_{T \to 0} \Delta S^{(\alpha)}$ decreases below $\log 2$ as shown in Fig.~\ref{fig:2ckresult}(d). Since $\Delta E$ is a relevant perturbation, we observe a non-monotonic temperature dependence for some range of values of $\Delta E$ as shown in  Fig.~\ref{fig:2ckresult}(d) (finite $\Delta E$).% Near zero temperature, $\Delta S_m^{(\alpha)}$ grows as the temperature increases, because the  critical state with $\Delta E=0$ is restored. But at some temperature of the order of $T_K$ the NEE starts to decrease as the temperature dephasing becomes stronger. We note that this nonmonotonic behavior does not take place for the RLM. For $J_-=0$, it occurs only at a large $\Delta E$ of the order of $T_K$. To judge whether the actual 2CK model shows this nonmonotonic behavior of the NEE one would need to study the NEE  directly from the model Eq.~(\ref{eq:MCK}) using methods like numerical renormalization group which we leave for future study.

%\paragraph*{Summary.}
To conclude, we %have 
proposed %in this paper 
an experimental procedure to measure entanglement at finite temperatures. In particular we %have 
demonstrated that an entanglement measure - the number entanglement entropy (NEE) -  can be obtained solely from the subsystems charge distribution function and correlation functions. 
We formulated exact thermodynamic relations between moments of the NEE and subsystem charge correlation functions. A similar observation was made for the quantum Fisher information% which witnesses multipartite entanglement
~\cite{hauke2016measuring}. %In our approach, the $\alpha$-th moment of the NEE of a thermal state at temperature $T$ is related to a charge correlation function of the same Hamiltonian at  temperature $ T/\alpha$. %From these relations, we can estimate the entanglement of  mixed states 
%without statistical averaging process~\cite{gray2018machine,zhou2020single,elben2020mixed}. 
The setup we propose  for measurement of the NEE in quantum dot systems has already been utilized to measure the charge noise, thus we expect that it can be readily extended to measure finite temperature entanglement measures in various systems,  including %the frustrated 
multi-channel Kondo systems.

%The number entanglement entropy (NEE), defined as the entropy change in a system with a conserved charge, due to a projective unselective measurement of a subsystem's charge, quantifies the bi-partite entanglement between the measured subsystem and its complement~\cite{macieszczak2019coherence,ma2021symmetric}. Here we addressed the question: to what extent can the NEE be obtained solely from the subsystems charge distribution function and correlation functions? We found that thermal states allow to formulate exact thermodynamic relations between moments of the NEE and subsystem charge correlation functions. A similar observation was made for the quantum Fisher information which witnesses multipartite entanglement~\cite{hauke2016measuring}. 

%In our approach, the $\alpha$-th moment of the NEE of a thermal state at temperature $T$ is 
%exactly related to a charge correlation function of the same Hamiltonian at  temperature $ T/\alpha$. We obtained explicit general relations for the first \Renyi moments of the NEE which can be applied in quantum dot systems. From these relations, we can estimate the entanglement of  mixed states 
%without statistical averaging process~\cite{gray2018machine,zhou2020single,elben2020mixed}. Using our approach we discussed a prospect for measurement of the number entanglement in quantum dots systems, including the frustrated multichannel-channel Kondo system. %The intriguing possibility to apply our relation to obtain the entanglement from time dependent subsystem charge measurements~\cite{kung2009noise} is left for a future study.

\begin{acknowledgments}
\paragraph*{Acknowledgments.} CH thanks Donghoon Kim for useful comments. We acknowledge support from the European Research
Council (ERC) under the European Unions Horizon
2020 research and innovation programme under grant
agreement No. 951541, ARO (W911NF-20-1-0013),
the US-Israel Binational Science Foundation (Grant No.
2016255) and the Israel Science Foundation, grant number 154/19. 
\end{acknowledgments}

\end{document}

\newcommand{\Tr}{\text{Tr}}
\title{Realistic protocol to measure entanglement at finite temperatures:\linebreak
Supplementary Material}
%\title{Thermodynamic relations between number entanglement  and charge correlations}
%Equilibrium relations between number entanglement and occupation noise

%Measuring\Renyi Entropy change using occupation noise
\author{Cheolhee Han}
\affiliation{School of Physics and Astronomy, Tel Aviv University, Tel Aviv 6997801, Israel}

\author{Yigal Meir}
\affiliation{Department of Physics, Ben-Gurion University of the Negev, Beer-Sheva, 84105 Israel}
\affiliation{Department of Physics, Princeton University, Princeton, NJ 08540}

\author{Eran Sela}
\affiliation{School of Physics and Astronomy, Tel Aviv University, Tel Aviv 6997801, Israel}

\maketitle
%\YM{I think we need to number the sections so we can refer to them in the main text}
\section{Positivity of $\Delta \smk$}
\label{se:appendixskmonotone}
In this section we % prove the positivity of the \Renyi moments $\Delta \smk \equiv \frac{1}{\ell-1}(\log(\Tr[\rho^\ell])-\log(\Tr[\rho_{\hat{N}_A}^\ell]))$ where $\rho_{\hat{N}_A} \equiv \sum_{N_A}\Pi_{N_A}\rho\Pi_{N_A}$ and $\Pi_{N_A}$ is a projector to a sector with fixed $N_A$.
%We 
would like to 
show that
\be
\label{eq:positivity}
-\log(\Tr[\rho_{\hat{N}_A}^\ell]) \ge -\log(\Tr[\rho^\ell]).
\ee
Since the $\ell$-th \Renyi entropy is Schur convex~\cite{bengtsson2006geometry}, this inequality is guaranteed if the eigenvalues of $\rho_{\hat{N}_A}$, denoted $\{p_i\}$ %($i=1,\dots N$) 
in nonincreasing order, %where $N$ is the dimension of the Hilbert space, 
are majorized by those of $\rho$ denoted $\{\lambda_i\}$, i.e. if
\be
   \vec{p}\prec \vec{\lambda}\label{eq:majorize0}.
\ee
%Explicitly, this condition reads $\sum_{i=1}^n p_i \le \sum_{i=1}^n \lambda_i$,  ($n=1,2,\dots,N$) and $\sum_{i=1}^N p_i = \sum_{i=1}^N \lambda_i$. 
Thus we just need to prove this majorization condition. By construction $\rho_{\hat{N}_A}$ is block diagonal in $N_A$. This means that we can diagonalize $\rho_{\hat{N}_A}$ by a unitary operator $U_{\hat{N}_A}$ which itself is block diagonal in $N_A$,
\begin{equation}
U_{\hat{N}_A}\rho_{\hat{N}_A} U_{\hat{N}_A}^\dagger={\rm{diag}}(\{p_i \}).
\end{equation}
This implies that the diagonal elements of the density matrix $U_{\hat{N}_A}\rho U_{\hat{N}_A}^\dagger$ are the same as those of $U_{\hat{N}_A}\rho_{\hat{N}_A} U_{\hat{N}_A}^\dagger$, i.e they are given by $\{p_i \}$. Applying the Schur-Horn theorem~\cite{bengtsson2006geometry} to the Hermitian matrix $U_{\hat{N}_A}\rho_{\hat{N}_A} U_{\hat{N}_A}^\dagger$, we conclude that its diagonal elements $\{p_i \}$ are majorized by its eigenvalues $\{\lambda_i \}$, i.e. Eq.~(\ref{eq:majorize0}) is satisfied, leading to Eq.~(\ref{eq:positivity}). Therefore $\Delta S^{(\ell)}$ is positive and zero when $\rho=\rho_{\hat{N}_A}$.

%By construction $\rho_m$ is block diagonal in $N_A$. This means that we can diagonalize $\rho_m$ by a unitary operator $U_m$ which itself is block diagoan as well,
%\begin{equation}
%U_m\rho_m U_m^\dagger=\rho_{m,diag}.
%\end{equation}
%This implies that the density matrix $U_m\rho U_m^\dagger$ has the same diagonal element as $\rho_{m,diag}$. Let us denote these diagonal elements by $p_i$ ($i=1,\dots n$) in nonincreasing order, where $n$ is the dimension of the Hilbert space. From the Schur-Horn theorem~\cite{bengtsson2006geometry}, for a Hermitian matrix $\rho$, its diagonal elements $\vec{p}$ is majorized by its eigenvalue vector \ES{shouldn't this be ``eigenvalues"?}  $\vec{\lambda}$, or
%\begin{align}
%    \vec{p}\prec \vec{\lambda}\label{eq:majorize}
%\end{align}
%Since the diagonal matrix $U_m\rho_m U_m^\dagger$ has the same diagonal elements with $U_m\rho U_m^\dagger$,  from Eq.~\eqref{eq:majorize}, the diagonal elements of the $U_m\rho U_m^\dagger$, or $U_m \rho_m U_m^\dagger$ is majorized by $\rho$. Since the $k$th \Renyi entropy is Schur convex~\cite{bengtsson2006geometry}, the inequality below is satisfied,
%\begin{equation}
%\log(\Tr[\rho^k])-\log(\Tr[\rho_m^k])\geq 0.
%\end{equation}
%So $\Delta S_m^{(k)}$ is positive and zero when $\rho=\rho_m$. 

\section{Derivation of Eq.~(7) in the main text}
\label{se:twochargestatecase}
%We now apply Eq.~(\ref{eq:central_general}) for the case where  subsystem $A$ consists of a quantum dot (QD) allowing only two charge states. Then the lowest moments of $\Delta \smk$ are attainable just in terms of  QD charge correlations in real time, at temperature $T/k$.

Consider a subsystem $A$ with only  two charge states, $N_A=0,1$. %This describes for example a QD which is either occupied or empty. %There is no restriction on subsystem $B$ which can be, e.g., another QD or a lead as we will discuss in the examples in Sec.~\ref{se:examples}. 
In this case the projection operators become $\Pi_{N_A=1}=\hat{N}_A$,  $\Pi_{N_A=0}=1-\hat{N}_A$. Focus on the first \Renyi moments. The trace of $\rho_{\hat{N}_A}^2$ and $\rho_{\hat{N}_A}^3$ is given by
\begin{align}
\Tr[\rho_{\hat{N}_A}^2]=&\Tr[\hat{N}_A\rho\hat{N}_A\rho]+\Tr[(1-\hat{N}_A)\rho(1-\hat{N}_A)\rho]\nonumber\\
=&\frac{1}{Z(T)^2}\Big[\sum_{i}e^{-2E_{i}/T}-2e^{-2E_i/T}\langle i|\hat{N}_A|i\rangle +2\sum_{ij}e^{-E_{i}/T-E_{j}/T}|\langle {i}|\hat{N}_A|{j}\rangle|^2 \Big], \label{eq:rhom2}
\end{align}
and
\begin{align}
\Tr[\rho_{\hat{N}_A}^3]=&\frac{1}{Z(T)^3}\Big[\sum_{i}e^{-3E_{i}/T}-3e^{-3E_i/T}\langle i|\hat{N}_A|i\rangle) +3\sum_{ij}e^{-E_{i}/T-2E_{j}/T}|\langle {i}|\hat{N}_A|{j}\rangle|^2 \Big],\label{eq:rhom3}
\end{align}
respectively. The first terms in Eqs.~\eqref{eq:rhom2} and \eqref{eq:rhom3} are the same as $Z(T/\ell)$ with $\ell=2,3$, respectively. Likewise the second terms involve the number occupation at temperature $T/\ell$,
\begin{align}
\sum_{i}\langle {i}|\hat{N}_A|{i}\rangle e^{-\ell E_{i}/T}=Z(T/\ell)\langle \hat{N}_A\rangle_{T/\ell}.\label{eq:num_kth}
\end{align}
%for $k=2$ and $3$. 
The third terms can be rewriten as
\begin{align}
&\sum_{ij}e^{-E_i/T}e^{-(\ell-1)E_j/T}|\langle i|\hat{N}_A|j\rangle|^2 =\sum_{ij}e^{-\ell E_i/T} e^{(\ell-1)(E_i-E_j)/T}|\langle i|\hat{N}_A|j\rangle|^2. \nonumber
\end{align}
This resembles an imaginary time charge correlator. To see this we define the imaginary time Green's function
\begin{align}
\langle \hat{N}_A(-i\tau)\hat{N}_A(0)\rangle_T =\sum_{ij}\frac{e^{-E_{i}/T}}{Z(T)}|\langle {i}|\hat{N}_A|{j}\rangle|^2e^{(E_{i}-E_{j})\tau},\label{eq:nncor_imag}
\end{align}
%Thus the third terms of Eq.~\eqref{eq:nncor_imag} can be written using
%\begin{equation}
%\begin{split}
%\Lambda\Big(\tau=\frac{1}{T},\frac{T}{2}\Big)=&\sum_{ij}\frac{e^{-(E_i+E_j)/T}}{Z(T/2)}|\langle i|\hat{N}_A|j\rangle|^2, \\
%\Lambda\Big(\tau=\frac{2}{T},\frac{T}{3}\Big)=&\sum_{ij}\frac{e^{-(E_i+2E_j)/T}}{Z(T/3)}|\langle i|\hat{N}_A|j\rangle|^2,
%\end{split}\label{eq:nncor_imag_half}
%\end{equation}
%respectively, yielding
and obtain
\begin{equation}
\begin{split}
\Tr[\rho_{\hat{N}_A}^2]=\frac{Z\big(\frac{T}{2}\big)}{Z(T)^2}\big[1-2\langle \hat{N}_A\rangle_{T/2}+2\langle \hat{N}_A(-i/T)\hat{N}_A(0)\rangle_{T/2}\big],\\
\Tr[\rho_{\hat{N}_A}^3]=\frac{Z\big(\frac{T}{3}\big)}{Z(T)^3}\big[1-3\langle \hat{N}_A\rangle_{T/3}+3\langle \hat{N}_A(-2i/T)\hat{N}_A(0)\rangle_{T/3}\big].
\end{split}
\end{equation}
Therefore the 2nd and 3rd \Renyi moments of the NEE  can be written in terms of imaginary time charge correlators as
\begin{equation}
\begin{split}
\Delta S^{(2)}(T)=&-\log\big[1-2\langle \hat{N}_A\rangle_{T/2}+2\langle \hat{N}_A(-i/T)\hat{N}_A(0)\rangle_{T/2}\big]
,\label{eq:main_result}\\
\Delta S^{(3)}(T)=&-\frac{1}{2}\log\big[1-3\langle \hat{N}_A\rangle_{T/3}+3\langle \hat{N}_A(-2i/T)\hat{N}_A(0)\rangle_{T/3}\big].
\end{split}
\end{equation}

At thermal equilibrium the imaginary time correlator can be related with the appropriate real time correlator. To see this, we write the real time correlator using the Lehmann representation,
\begin{align}
&\chi(\omega,T)=\int dt e^{i\omega t}\langle \hat{N}_A(t)\hat{N}_A(0)\rangle_{T}=\frac{1}{Z(T)}\sum_{ij}|\langle {i}|\hat{N}_A|{j}\rangle|^2 e^{- E_{i}/T}2\pi \delta(\omega+E_{i}-E_{j}).\label{eq:nncor_freqsym_Lehmann}
\end{align}
Comparing with the imaginary time Green's function in Eq.~\eqref{eq:nncor_imag}, we obtain the relation
\begin{align}
\langle \hat{N}_A(-i\tau)\hat{N}_A(0)\rangle_T=\int \frac{d\omega}{2\pi} \chi(\omega,T) e^{-\omega\tau}.\label{eq:nncor_RI_rel}
\end{align}
%Therefore $\Lambda(\tau,T)$ is obtained from $\chi(\omega,T)$. After putting specific $\tau$ and $T$ as in Eq.~\eqref{eq:nncor_imag_half}, we obtain the change of the 2nd and 3rd RE %shown in Eqs.~\eqref{eq:central_Renyi23}. Using this way, in principle we can relate $k$-point imaginary time function (see Eq.~\eqref{eq:central_general}) and $k$-point real time correlator. 
As a result, Eq.~(\ref{eq:main_result}) becomes Eq.~(7) in the main text.%, which we rewrite here for clarity,
%\begin{equation}
%\begin{split}
%\Delta S^{(2)}=&-\log\Big[1-2\langle\hat{N}_A\rangle_{T/2}+2\int\frac{d\omega}{2\pi}\chi\big(\omega,T/2\big) e^{-\frac{\omega}{T}}\Big],\\
%\Delta S^{(3)}=&-\frac{1}{2}\log\Big[1-3\langle \hat{N}_A\rangle_{T/3}+3\int \frac{d\omega}{2\pi}\chi\big(\omega,T/3)e^{-\frac{2\omega}{T}}\Big].
%\end{split}\label{eq:central_Renyi23SM}
%\end{equation}

%where $\chi(\omega,T)$ is the charge noise at temperature $T$,
%\begin{align}
%\chi(\omega,T)\equiv\int dt e^{i\omega t} \langle \hat{N}_A(t)\hat{N}_A(0)\rangle_T.\label{eq:nncor_freqsym}
%\end{align}
%This is our main result. It allows one to measure \Renyi moments of the NEE in thermal states at temperature $T$, from charge correlations measured at temperature $T/k$. Next, we discuss how $\chi(\omega)$ can be actually obtained in mesoscopic systems.

%More generally what we achieved in this section is to relate the $k$-point imaginary time function (see Eq.~\eqref{eq:central_general}) and $k$-point real time correlator. 

\section{Charge-charge correlator from shot noise via a QPC}
%We envision that a charge detector attached to QD $A$ will measure both the average charge and its correlation function at different temperatures $\chi(\omega,T)$. Whereas average charge measurements are standard, a measurement of the QD charge correlation $\chi(\omega)$ was not achieved so far to the best of our knowledge. {\color{red}Yet, its feasibility seems reasonable given a large body of work on noise of mesoscopic systems~\cite{blanter2000shot}.}
In this section we provide a
simple  model in which $\chi(\omega)$ can be related to the voltage-dependent noise in the %current of the
charge detector quantum point contact (QPC). 

 \begin{figure}[]
	\includegraphics[width=0.5\columnwidth]{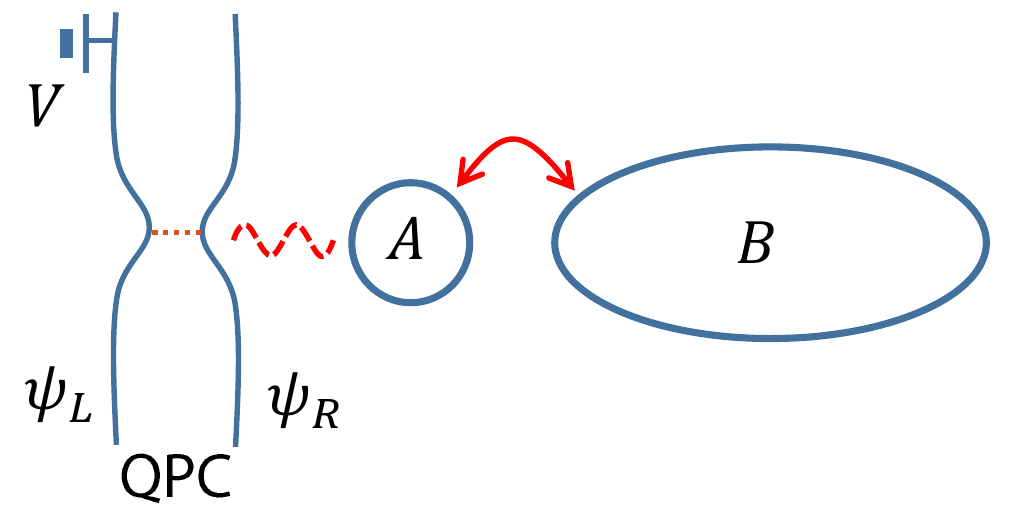}
	\caption{A QPC interacts electrostatically with subsystem $A$. The backscattering strength between modes $\psi_L$ and $\psi_R$ in the QPC depends on $N_A$. %The left mode is biased by a voltage $V$ while the right mode is grounded. 
	}\label{fig:expsetup}
\end{figure}

Consider subsystem $A$ to describe  a QD whose charge is monitored using a QPC, see Fig.~\ref{fig:expsetup}. 
We assume the QPC  consists of one left moving and one right moving noninteracting fermion mode, $\psi_L$ and $\psi_R$, undergoing a weak backscattering (see dashed line in Fig.~\ref{fig:expsetup}) at the QPC. The backscattering amplitude consists of two terms, one which depends on the charge of QD $A$ with coefficient $\lambda$, and one which is a constant $\lambda_0$,
\begin{equation}
\begin{split}
H_{T}=& \lambda \hat{N}_A \psi^\dagger_L(t)\psi_R(t)+\text{h.c.},\\
H_{T_0}=&\lambda_0 \psi^\dagger_L(t)\psi_R(t)+\text{h.c.}.
\end{split}
\end{equation}
We set the voltage bias of the left moving mode to be $V$ and ground the right moving mode. %The tunneling Hamiltonian of the QPC is
The current operators of the QPC are $\hat{I}_{T_{(0)}}=i[N_L-N_R,H_{T_{(0)}}]$, where $N_{L(R)}$ is the particle number of the left (right) mode.
%Measurement of the 2nd and 3rd moments of the NEE, Eq.~(7) in the main text requires to extract $\langle\hat{N}_A\rangle$ and $\chi(\omega)$ at temperature $T/k$. The former is extracted from the charge dependent current $\langle I_{T}\rangle$.

We now demonstrate how $\chi(\omega)$ can be obtained. Setting $\lambda_0=0$ with finite $\lambda\ne 0$, we consider the current $I(V,T)$ and the symmetrized noise $S(V,T)=\int dt [\frac{1}{2}\langle \{\hat{I}_T(t),\hat{I}_T(0)\}\rangle -\langle \hat{I}_T\rangle^2 ]$. 
After a derivation presented in %the Appendix~\cite{SM}, 
the next subsection, we obtain
%To extract $\chi(\omega,T)$, we  Fourier transform $f(V,T)$, \begin{align}&\int \frac{dV}{2\pi} f(V,T)e^{iV y}=\frac{\pi T^2  y^2}{\sinh^2(\pi T y)}\int \frac{d\omega}{2\pi} \chi(\omega,T)e^{i\omega y}.\end{align}
%$\chi(\omega,T)$ as
\begin{align}
\chi(\omega,T)=\int dy \int \frac{dV}{2\pi}&\partial_V^2\Big[\frac{S(V,T)+e I(V,T)}{2\lambda^2 e^2}\Big]\frac{\sinh^2(\pi Ty)}{\pi T^2 y^2}e^{i(V-\omega)y}.\label{eq:numcor_rel}
\end{align}
This relation allows to extract the charge-correlation from the finite voltage noise and current at the QPC. %{\color{red}Given the well developed techniques of charge and noise measurements~\cite{blanter2000shot} we believe that measuring $S_m^{(2,3)}$ is feasible.}

\subsection{Derivation of Eq.~(\ref{eq:numcor_rel})% in the main text
}
\label{se:appendixnoisechi}
In this subsection, we  derive Eq.~(\ref{eq:numcor_rel}) %Eq.~(9) in the main text 
using perturbation theory in $\lambda$. First we calculate the noise up to the lowest order of $\lambda$,
\begin{align}
S(V,T)=\frac{\lambda^2e^2}{2}\int_{-\infty}^{\infty}dt&\Big[\langle \psi_L^\dagger(t)\psi_R(t)\psi^\dagger_R(0)\psi_L(0)\rangle\langle \hat{N}_A(t)\hat{N}_A(0)\rangle +\langle\psi_R^\dagger(t)\psi_L(t)\psi_L^\dagger(0)\psi_R(0)\rangle \langle \hat{N}_A(t)\hat{N}_A(0)\rangle \nonumber\\
&+\langle \psi_L^\dagger(0)\psi_R(0)\psi^\dagger_R(t)\psi_L(t)\rangle\langle \hat{N}_A(0)\hat{N}_A(t)\rangle +\langle \psi_R^\dagger(0)\psi_L(0)\psi^\dagger_L(t)\psi_R(t)\rangle\langle \hat{N}_A(0)\hat{N}_A(t)\rangle \Big].\label{eq:supp_noise1}
\end{align}
Next, using $\langle \hat{N}_A(t)\hat{N}_A(0)\rangle=\langle \hat{N}_A(0)\hat{N}_A(-t)\rangle$, and the correlators for $\psi_L$ and $\psi_R$,
\begin{align}
\langle \psi_L^\dagger(t) \psi_L(0)\rangle=\frac{e^{iVt}}{\frac{\beta}{\pi}\sin\big(\frac{\pi}{\beta}(\tau_0+it)\big)},\quad \langle \psi_R^\dagger(t)\psi_R(0)\rangle=\frac{1}{\frac{\beta}{\pi}\sin\big(\frac{\pi}{\beta}(\tau_0+it)\big)},
\end{align}
where $\tau_0\equiv a_0/v_F$ with infinitesimal length $a_0$, we obtain 
\begin{align}
S(V,T)=&\lambda^2 e^2\int_{-\infty}^{\infty} dt \frac{1}{\Big[\frac{\beta}{\pi}\sin(\frac{\pi}{\beta}(\tau_0+it))\Big]^2}(e^{iVt}+e^{-iVt})\langle \hat{N}_A(t)\hat{N}_A(0)\rangle. \label{eq:noise_supp}
\end{align}
Repeating the same algebra for the current, we obtain
\begin{align}
I(V,T)=\lambda^2 e\int_{-\infty}^0 dt&\Big[\langle \psi_L^\dagger(t) \psi_R(t)\psi^\dagger_R(0)\psi_L(0)\rangle\langle N_A(t)N_A(0)\rangle -\langle \psi_R^\dagger(t)\psi_L(t)\psi_L^\dagger(0)\psi_R(0)\rangle\langle N_A(t)N_A(0)\rangle \nonumber\\
&-\langle \psi_R^\dagger(0)\psi_L(0)\psi_L^\dagger(t)\psi_R(t)\rangle\langle N_A(0)N_A(t)\rangle +\langle \psi_L^\dagger(0)\psi_R(0)\psi_R^\dagger(t)\psi_L(t)\rangle\langle N_A(0)N_A(t)\rangle \Big]\nonumber\\
=&\lambda^2 e\int_{-\infty}^{\infty}dt \frac{1}{\Big[\frac{\beta}{\pi}\sin(\frac{\pi}{\beta}(\tau_0+it))\Big]^2}(e^{iVt}-e^{-iVt})\langle N_A(t)N_A(0)\rangle.\label{eq:cur_supp}
\end{align}
Combining the two observables,
\begin{align*}
\frac{S(V,T)+e I(V,T)}{2\lambda^2 e^2}=\int_{-\infty}^\infty dt \frac{e^{iV t}}{\Big[\frac{\beta}{\pi}\sin(\frac{\pi}{\beta}(\tau_0+it))\Big]^2}\langle N_A(t)N_A(0)\rangle.
\end{align*}
Fourier transforming, we find
\begin{align}
\frac{S(V,T)+eI(V,T)}{2\lambda^2 e^2}&=\int\frac{d\omega}{2\pi}\chi(\omega,T)(\omega-V)\Big[1+\coth(\frac{\omega-V}{2T})\Big].\label{eq:supp_numcor_exp1}
\end{align}
After taking the second derivative with respect to $V$, we have
\be
\partial_V^2\Big[\frac{S(V,T)+eI(V,T)}{2\lambda^2 e^2}\Big] =\int\frac{d\omega}{2\pi}\chi(\omega,T)\frac{\frac{\omega-V}{2T}\coth\Big(\frac{\omega-V}{2T}\Big)-1}{T\sinh^2(\frac{\omega-V}{2T})}.\label{eq:numcor_exp}
\ee
To extract $\chi(\omega,T)$, we  Fourier transform the quantity, 
\begin{align}
&\int \frac{dV}{2\pi} \partial_V^2\Big[\frac{S(V,T)+eI(V,T)}{2\lambda^2 e^2}\Big] e^{iV y}=\frac{\pi T^2  y^2}{\sinh^2(\pi T y)}\int \frac{d\omega}{2\pi} \chi(\omega,T)e^{i\omega y}.
\end{align}
Finally, to obtain Eq.~(\ref{eq:numcor_rel}) %the Eq.~(9) in the main text, 
we use the integral
\begin{align}
\int_{-\infty}^\infty \frac{dV}{2\pi} \frac{\frac{V}{2T}\coth\Big(\frac{V}{2T}\Big)-1}{T\sinh^2(\frac{V}{2T})} e^{i V y}=\frac{\pi T^2 y^2}{\sinh^2(\pi T y)}.
\end{align}

\begin{figure}[t]
\includegraphics[width=0.6\columnwidth]{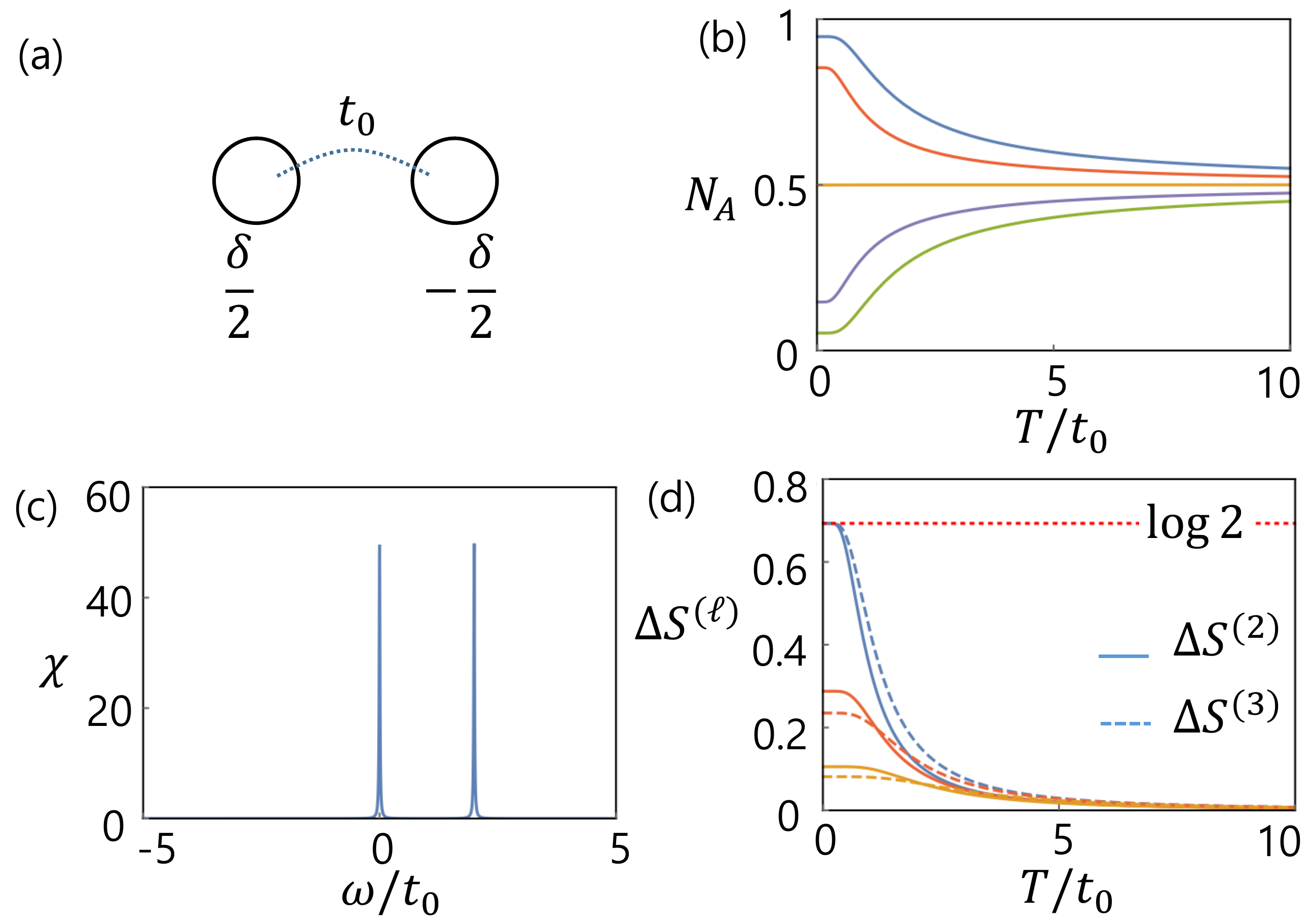}
\caption{(a) Double dot setup, see Eq.~\eqref{eq:DDham}. (b) $\langle \hat{N}_A\rangle_{T}$ versus $T$ for $t_0=1$ and $\delta=-2,-1,0,1,2$ from uppermost to lowermost. (c) $\chi(\omega,T)$ for $T=0.1$, $t_0=1$ and $\delta=0$. Two peaks correspond to $\omega=0$, $2\Delta$ (an additional peak at $\omega=-2\Delta$ is suppressed for $T \ll \Delta$. 
 (d) $\Delta S^{(2)}$ (solid lines) and $\Delta S^{(3)}$ (dashed lines) versus $T$ for $\delta=0,1,2$ from uppermost to lowermost with $t_0=1$. }\label{fig:ddresult}
\end{figure}
\section{Simplest example : Double dot }
In this section we exemplify Eq.~(7) of the main text for a double quantum dot. We compute the subsystem charge and its correlation, from which the \Renyi moments of the NEE are obtained.

Consider two QDs denoted $A$ and $B$ with Hamiltonian 
\begin{align}
H=t_0 (c_A^\dagger c_B+c_B^\dagger c_A)+\delta(c_A^\dagger c_A-c_B^\dagger c_B),\label{eq:DDham}
\end{align}
where $c_{A(B)}^\dagger$ creates a spinless electron in QD $A(B)$, $t_0$ is the hopping amplitude, $2\delta$ is the asymmetry of the QDs, and we assume the double dot contains one electron, see Fig.~\ref{fig:ddresult}(a).

At zero temperature, the NEE moments $\Delta S^{(2)}$ and $\Delta S^{(3)}$ approach $\log(2)$ in the symmetric  case $\delta\to 0$ where the wave function becomes a Bell state $(|10\rangle + |01\rangle)/\sqrt{2}$ where $1(0)$ denotes an occupied (empty) QD. They remain finite at any $T>0$ and decay as~\cite{ma2021symmetric} $1/T^2$ as shown in Fig.~\ref{fig:ddresult}(d) for both $\Delta S^{(2)}$ and $\Delta S^{(3)}$. 

We now  calculate separately the ingredients of Eq.~(7) of the main text. The average and two point correlator of $N_A$ are given by
\be
\langle \hat{N}_A\rangle_{T}=a^2 f(-\Delta/T)+b^2 f(\Delta/T),\label{eq:DDnum}
\ee
and 
\be
\frac{\chi(\omega,T)}{2\pi}=a^2 b^2\Big(\delta(\omega+2\Delta)  f(\frac{\Delta}{T})^2+\delta(\omega-2\Delta) f(-\frac{\Delta}{T})^2\Big)
+\delta(\omega)\Big[ a^4 f(-\frac{\Delta}{T})+b^4 f(\frac{\Delta}{T})+ 2a^2b^2 f(\frac{\Delta}{T})f(-\frac{\Delta}{T})\Big] ,\label{eq:DDnoise}
\ee
respectively, where  $a=\sin\alpha/2$, $b=\cos\alpha/2$ with $\alpha=\tan^{-1}t_0/\delta$,  $\Delta=\sqrt{t_0^2+\delta^2}$, and $f(x)=1/(1+e^x)$. They are plotted in Fig.~\ref{fig:ddresult}(b) and (c). In the symmetric case $\langle N_A \rangle = 1/2$ and the information about the NEE is encoded in the two point charge correlator $\chi(\omega)$, which in this system features $\delta$-function peaks at the eigenenergy differences $0,\pm 2 \Delta$ with appropriate matrix elements. 

%We can also infer the imaginary time correlation function $\Lambda$ by substituting  Eq.~\eqref{eq:DDnoise} into Eqs.~\eqref{eq:nncor_RI_rel}, yielding \begin{align}\Lambda(\tau,T)=\frac{1+a^4e^{\Delta/T}+b^4e^{-\Delta/T}+2a^2b^2\cosh(\frac{\Delta}{T}-2\Delta\tau)}{4\cosh^2(\frac{\Delta}{2T})},& \nonumber\\&\label{eq:DDcor_imag}\end{align}as plotted in Fig.~\ref{fig:ddresult}(c). 

Substituting Eqs.~(\ref{eq:DDnum}) and~(\ref{eq:DDnoise}) into Eq.~(7) of the main text, one obtains 
\bea
\Delta S^{(2)}&=&-\log\Big[\frac{1+\cosh(\frac{2\Delta}{T})(a^4+b^4)+2a^2b^2}{2\cosh^2(\frac{\Delta}{T})}\Big], \nonumber \\
\Delta S^{(3)}&=&-\frac{1}{2}\log\Big[\frac{1}{2\cosh^2(3\Delta/2T)}\Big(1+(a^6+b^6)\cosh(\frac{3\Delta}{T})+3(a^4 b^2+a^2b^4)\cosh(\frac{\Delta}{T})\Big)\Big],\label{eq:DD_Renyi3}
\eea
which  can be derived directly from the density matrix. 

%In this simple example one can in principle extract any measure of entanglement, such as as the von-Neumann NE, using full state tomography. We now move to a more complex many-body example where this is no longer possible.

\subsection{Derivation of Eq.~(\ref{eq:DDnoise})}
First we diagonalize the double dot Hamiltonian as
\begin{align}
H=\Delta c_1^\dagger c_1-\Delta c_0^\dagger c_0,
\end{align}
where $c_A=a c_0+b c_1$, $c_B=bc_0-ac_1$. Then the charge correlator becomes
\begin{align}
\langle c_A^\dagger c_A(t) c_A^\dagger c_A(0)\rangle=&a^4\langle c_0^\dagger c_0(t)c_0^\dagger c_0(0)\rangle+b^4\langle c_1^\dagger c_1(t)c_1^\dagger c_1(0)\rangle\nonumber\\
&+a^2 b^2\big(\langle c_0^\dagger c_0(t) c_1^\dagger c_1(0)\rangle+\langle c_1^\dagger c_1(t) c_0^\dagger c_0(0)\rangle +\langle c_1^\dagger c_0(t) c_0^\dagger c_1(0)\rangle+\langle c_0^\dagger c_1(t) c_1^\dagger c_0(0)\rangle\big)\nonumber\\
=&\frac{a^4}{(1+e^{-\Delta/T})}+\frac{b^4}{1+e^{\Delta/T}}+\frac{2a^2 b^2}{(1+e^{\Delta/T})(1+e^{-\Delta/T})}+\frac{a^2 b^2 e^{2i\Delta t}}{(1+e^{\Delta/T})^2}+\frac{a^2 b^2 e^{-2i\Delta t}}{(1+e^{-\Delta/T})^2}.\label{eq:supp_dd}
\end{align}
Eq.~(\ref{eq:DDnoise}) follows by Fourier transforming Eq.~\eqref{eq:supp_dd}.

\section{Derivation of 2-channel Kondo results}
In this section we detail the derivation of results for the number operator, charge noise, as well as \Renyi entropies for the 2CK model based on the effective non-interacting Majorana fermion theory. We also explain our novel numerical method to obtain the NEE $\Delta S$ which was not obtained on  previous calculations which considered exclusively the \Renyi moments of the NEE~\cite{ma2021symmetric}.

\subsection{\Renyi entropies}
%\ES{Cheolhee, can you give a specific reference with the closest notation to the one you use?}
First we rewrite the Hamiltonian~\cite{schiller1998toulouse} 
\begin{align}
H=\sum_k \frac{k}{2}\gamma_{-k}\gamma_k +\frac{k}{2}\eta_{-k}\eta_k +i\sqrt{\frac{2T_K}{L}}\sum_{k}\gamma_{k}\eta_d +i\sqrt{\frac{2T_-}{L}}\sum_k \eta_k\gamma_d +ih\gamma_d\eta_d.
\end{align}
The Majorana mode operators $\gamma_k$ and $\eta_k$ satisfy $\{\gamma_k,\gamma_q\}=\delta_{q,-k}$, $\{\eta_k,\eta_q\}=\delta_{q,-k}$, and $\{\gamma_k,\eta_q\}=0$, and the local Majorana zero modes satisfy $\{\gamma_d,\gamma_d\}=\{\eta_d,\eta_d\}=1$.
%\YM{we need to define these operators}
The number operator is
\begin{align}
\hat{N}_d=c_d^\dagger c_d=i\gamma_d\eta_d+1/2
\end{align}
Since the Hamiltonian is non-interacting, we use Wick's theorem to obtain
\begin{align}
\chi(\omega,T)=\int dt e^{i\omega t}\langle \hat{N}_d(t)\hat{N}_d(0)\rangle=&-2\pi\delta(\omega)\Big[\int \frac{d\omega_1}{2\pi}\langle \gamma_d\eta_d\rangle(\omega_1)\Big]^2+2\pi \delta(\omega) i\int \frac{d\omega_1}{2\pi}\langle \gamma_d\eta_d\rangle(\omega_1)+\frac{\pi}{2}\delta(\omega)\nonumber\\
&+\int\frac{d\omega_1}{2\pi}\langle\gamma_d\gamma_d\rangle(\omega_1)\langle\eta_d\eta_d\rangle(\omega-\omega_1)-\int \frac{d\omega_1}{2\pi}\langle \gamma_d\eta_d\rangle(\omega_1)\langle \eta_d\gamma_d\rangle(\omega-\omega_1).
\end{align}
The Green functions here are
\begin{equation}
    \begin{split}
        \langle \gamma_d\gamma_d\rangle(\omega)=\frac{2T_-\omega^2+2 T_- T_K^2+2T_K h^2}{(\omega^2-T_K T_- -h^2)^2+(T_K+T_-)^2\omega^2}\frac{1}{1+e^{-\beta\omega}},\\
\langle \eta_d\eta_d\rangle(\omega)=\frac{2 T_K\omega^2+2 T_K T_-^2+2T_- h^2}{(\omega^2-T_K T_- -h^2)^2+(T_K+T_-)^2\omega^2}\frac{1}{1+e^{-\beta \omega}},\\
\langle \eta_d\gamma_d\rangle(\omega)=\frac{2 h(T_K+T_-)\omega}{(\omega^2-T_K T_--h^2)^2+(T_K+T_-)^2\omega^2}\frac{1}{1+e^{-\beta \omega}},\\
\langle \gamma_d\eta_d\rangle(\omega)=-\frac{2 h(T_K+T_-)\omega}{(\omega^2-T_K T_- -h^2)^2+(T_K+T_-)^2\omega^2}\frac{1}{1+e^{\beta \omega}}.
    \end{split}
\end{equation}
Using these results we obtain $\chi(\omega)$ in Fig.~1(c) in the main text. Similarly the results in Fig.~1(d-f) in the main text are obtained using Eq.~(7) in the main text.

\subsection{Gaussian property of blocks in $\rho_{\hat{N}_A}$}
In this subsection we provide a general approach to compute the NEE $\Delta S$ for  free fermion models with a subsystem consisting of a single site, using Gaussian properties of the measured density matrix. This method was used for the results displayed in Fig.~1(d,e,f) of the main text.

Consider the density matrix of $N+1$ sites. After measuring one site ($A$ site), the measured density matrix is
\begin{align}
    \rho_{\hat{N}_A}=\hat{N}_A\rho \hat{N}_A + (1-\hat{N}_A)\rho(1-\hat{N}_A).
\end{align}
Now we show that $\hat{N}_A\rho \hat{N}_A$ or $(1-\hat{N}_A)\rho(1-\hat{N}_A)$ have a Gaussian form. We write the projectors $\hat{N}_A=c_A^\dagger c_A$ and $1-\hat{N}_A$  as
\begin{align}
    \hat{N}_A=\lim_{\alpha\rightarrow\infty} e^{\alpha (c_A^\dagger c_A-1)},\quad 1-\hat{N}_A=\lim_{\alpha\rightarrow\infty}e^{-\alpha c_A^\dagger c_A},
\end{align}
which is of course Gaussian. Since the product of Gaussian matrices is Gaussian, $\hat{N}_A\rho\hat{N}_A$ and $(1-\hat{N}_A)\rho(1-\hat{N}_A)$ are Gaussian. 
We define new density matrices as
\begin{align}
    \rho^{N_A=1}=\frac{1}{\langle \hat{N}_A\rangle}\hat{N}_A\rho \hat{N}_A,\quad \rho^{N_A=0}=\frac{1}{1-\langle \hat{N}_A\rangle}(1-\hat{N}_A)\rho(1-\hat{N}_A),
\end{align}
which can be written in diagonal Gaussian forms
\begin{align}
    \rho^{N_A=1}=\frac{1}{K_{1}}e^{-\sum_q \epsilon_q^{N_A=1} a_q^\dagger a_q},\quad \rho^{N_A=0}=\frac{1}{K_0}e^{-\sum_q\epsilon_q^{N_A=0} b_q^\dagger b_q},
\end{align}
where $a_k$, $b_k$ and $c_i$ are related as
\begin{align}
    c_i=\sum_k \phi_i^{N_A=1}(k) a_k,\quad c_i=\sum_k \phi_i^{N_A=0}(k)b_k.
\end{align}
Then the entropy of those density matrices is~\cite{peschel2009reduced}
\begin{align}
    S_{N_A}=-\Tr[\rho^{N_A}\log\rho^{N_A}]=\sum_{l}\log(1+e^{-\epsilon_l^{N_A}})+\sum_{l}\frac{\epsilon_l^{N_A}}{e^{\epsilon_l^{N_A}}+1}.
\end{align}
The energy levels are obtained from the correlation matrix, 
\begin{align}
    C^{N_A=1}_{ij}=\frac{1}{\langle \hat{N}_A\rangle}\Tr[\hat{N}_A\rho \hat{N}_A c_i^\dagger c_j],\quad C^{N_A=0}_{ij}=\frac{1}{1-\langle \hat{N}_A\rangle}\Tr[(1-\hat{N}_A)\rho(1-\hat{N}_A)c_i^\dagger c_j],
\end{align}
where $i,j\ne A$. If $i$ or $j$ coincide with the site corresponding to subsystem $A$, then the correlator vanishes. 

After diagonalizing the correlation matrix, the correlator becomes
\begin{align}
    C^{N_A}_{ij}=\sum_{k}\phi_i^{N_A}(k)\phi^{N_A}_j(k)\frac{1}{e^{\epsilon^{N_A}_k}+1}\rightarrow \bar{C}^{N_A}_k=\frac{1}{e^{\epsilon_k^{N_A}}+1}.
\end{align}
If we diagonalize the correlation matrix, then  $\epsilon_k$ becomes
\begin{align}
    \epsilon_k^{N_A}=\log(\frac{1-\bar{C}^{N_A}_k}{\bar{C}^{N_A}_k}).
\end{align}

Finally we obtain 
\begin{align}
    \Delta S=\Tr[\rho\log\rho]+\langle N_A\rangle S_{N_A=1}+(1-\langle N_A\rangle)S_{N_A=0}-\langle N_A\rangle\log\langle N_A\rangle-(1-\langle N_A\rangle)\log(1-\langle N_A\rangle).
\end{align}
This allowed us to obtain the results plotted in Fig.~1(d,e,f) in the main text. In these calculations we use the lattice model
\begin{align}
    H=\Delta_L\sum_{n=-N_{max}-1/2}^{N_{max}+1/2} n\gamma_{-n}\gamma_n +n\eta_{-n}\eta_n+i\sqrt{\frac{2T_K}{L}}\sum_n \gamma_n \eta_d+i\sqrt{\frac{2T_-}{L}}\sum_n \eta_n \gamma_d+i\Delta E \gamma_d\eta_d,
\end{align}
with $\Delta_L=1$, $N_{max}=800$, $T_K=25\pi$.
%This relation satisfies both charge conserving and charge not conserving Hamiltonians. 

\section{Scaling of NEE in the multi-channel Kondo effect}
In this section we show for the MCK model that the low temperature scaling of $\Delta S^{(\ell)}$ is independent on $\ell=1,2,3...$ and is determined by the scaling dimension of the spin operator $S_z$.

For the MCK model the projection operators for the impurity spin are
\begin{align}
    \hat{N}_d=\frac{1}{2}+S_z,\quad 1-\hat{N}_d=\frac{1}{2}-S_z\label{eq:N_Srel}.
\end{align}
Consider first $\Delta S^{(2)}$. It can be written as
\be
\label{deltaSmCORRELATOR}
    \Delta S^{(2)}=\log(\Tr[\rho^2])-\log(\Tr[\rho \hat{N}_d\rho \hat{N}_d]+\Tr[\rho(1-\hat{N}_d)\rho(1-\hat{N}_d)])=-\log\Big[\frac{1}{2}+2\langle S_z(\frac{1}{iT})S_z(0)\rangle_{T/2}\Big].
\ee
We can expand the impurity spin operator as a sum over local fields of increasing scaling dimension,
%\begin{align}
%    S_z=\sum_i c_i\psi_i.
%\end{align}
%Here $c_i$ are constants, and  $\psi_i$ is a list of local field operators. $\psi_1$ is the leading operator which has a known scaling dimension $\Delta_M$ for $M$-channel Kondo effect. Below we  keep only the leading operator and write 
$S_z=c_1\psi+... $ where $\psi$ is the leading operator which has a known scaling dimension $\Delta_M$ and $...$ corresponds to operators with higher scaling dimensions. This operator determines the leading  low temperature behaviour of the spin-spin correlator in Eq.~(\ref{deltaSmCORRELATOR}), which becomes
\begin{align}
\langle S_z(\frac{1}{iT})S_z(0)\rangle\simeq c_1^2\Big[\frac{\pi T/2}{ \sin(\pi T /2(1/T))} \Big]^{2\Delta_{M}}\propto T^{2\Delta_{M}}.
\end{align}
Thus we obtain
\begin{align}
    \Delta S^{(2)}\simeq -\log(1/2+a T^{2\Delta_{M}})\simeq \log 2 -2 a T^{2\Delta_{M}}.
\end{align}
It is easy to extend this result to $\Delta S^{(\ell)}$ with $\ell>2$, and show that it satisfies the same temperature scaling. % Consider $\ell>2$. 
From Eq.~(5) in the main text, and Eq.~\eqref{eq:N_Srel}, we obtain
\begin{align}
    \Delta S^{(\ell)}=\frac{1}{1-\ell}\log\bigg[&\bigg\langle \Big[\frac{1}{2}+S_z\Big(\frac{\ell-1}{iT}\Big)\Big]\Big[\frac{1}{2}+S_z\Big(\frac{\ell-2}{iT}\Big)\Big]\cdots\Big[\frac{1}{2}+S_z\Big(\frac{1}{iT}\Big)\Big]\Big[\frac{1}{2}+S_z(0)\Big]\bigg\rangle_{T/\ell}\nonumber\\
    &+\Big\langle\Big[\frac{1}{2}-S_z\Big(\frac{\ell-1}{iT}\Big)\Big]\Big[\frac{1}{2}-S_z\Big(\frac{\ell-1}{iT}\Big)\Big]\cdots \Big[\frac{1}{2}-S_z\Big(\frac{1}{iT}\Big)\Big]\Big[\frac{1}{2}-S_z(0)\Big]\Big\rangle_{T/\ell} \bigg].
\end{align}
After expanding we find
\begin{align}
    \Delta S^{(\ell)}=\frac{1}{1-\ell}\log\big[\frac{1}{2^{\ell}}+2\sum_{k=1}^{\ell-1}\sum_{q=0}^{k-1}\langle S_z(i k/T)S_z(i q/T)\rangle_{T/\ell}+\cdots\big],
\end{align}
where $\cdots$ includes higher (4, 6 ...) point correlators, which have the temperature scaling dependence higher than $2\Delta_M$. Hence they are negligible at the low temperature limit. The sum of the two-point correlators gives the same temperature scaling.

Next we consider the NEE $\Delta S^{(1)}$. It can be written as
\begin{align}
    \Delta S^{(1)}=\Tr[\rho\log\rho]-\Tr[\rho (\frac{1}{2}+S_z) \log(\rho (\frac{1}{2}+S_z))]-\Tr[\rho (\frac{1}{2}-S_z)\log(\rho(\frac{1}{2}-S_z))].
\end{align}
Expanding the logarithm using 
\begin{align}
    \log A=-\sum_{k=1}^\infty\frac{1}{k}(I-A)^k,
\end{align}
the NEE becomes
\begin{align}
    \Delta S^{(1)}=&\frac{1}{Z(T)}\sum_{k=1}^\infty\frac{1}{k}\Big[-\Tr[e^{-H/T}(I-e^{-H/T})^k]+\Tr[e^{-H/T} (\frac{1}{2}+S_z)(I-e^{-H/T} (\frac{1}{2}+S_z))^k]\nonumber\\
&\qquad\qquad+\Tr[e^{-H/T} (\frac{1}{2}-S_z)(I-e^{- H/T}(\frac{1}{2}-S_z))^k]\Big].
\end{align}
Now we expand the last expression. From each factor we can choose either the $1/2$ or the $S_z$. First we consider the term with all the $1/2$'s chosen. It is equivalent to
\begin{align}
    \frac{1}{Z(T)}\Big[\Tr[e^{-H/T}\log(e^{- H/T} )]-\frac{1}{2}\Tr[e^{-H/T}\log (e^{- H/T} \frac{1}{2})]-\frac{1}{2} \Tr[e^{-H/T}\log (e^{-H/T}\frac{1}{2})]\Big]=\log(2).
\end{align}
Next we consider terms involving the $S_z$'s. At the low temperature, the dominant contribution comes two-point functions of $S_z$,
\begin{align}
    \sum_{k=1}^\infty \frac{1}{k} \sum_q^k \frac{k!}{(k-q)! q!} \frac{1}{2^{q-1}}\frac{Z(T/q)}{Z(T)}\sum_p^q q\langle S_z(\frac{p}{iT})S_z(0)\rangle_{T/q},
\end{align}
with
\begin{align}
    \langle S_z(\frac{p}{iT})S_z(0)\rangle_{T/q}=\Big[\frac{\pi T/q}{\sin(\pi p/q )}\Big]^{2\Delta_M}\sim T^{2\Delta_M}.
\end{align}
Therefore the leading low temperature correction of $\Delta S^{(1)}$ coincides with that of $\Delta S^{(\ell)}$ with $\ell=2,3...$.